\documentclass[twocolumn,showpacs,preprintnumbers,amsmath,amssymb,superscriptaddress]{revtex4-1}

\usepackage{graphicx}
\usepackage{dcolumn}
\usepackage{bm}
\usepackage{tabularx}

\usepackage{color}   
\usepackage{hyperref}
\hypersetup{
    colorlinks=true, 
    linktoc=all,     
    linkcolor=blue,  
}

\makeatletter
\newcommand{\colorcaption}[2][]{%
  \begingroup%
  \renewcommand{\@caption@fignum@sep}{ (Color online). }%
  \caption[#1]{#2}%
  \endgroup%
}
\makeatother

\begin{document}

\preprint{}

\title{Isospin symmetry in $B(E2)$ values: Coulomb excitation study of ${}^{21}$Mg}

\author{P.~Ruotsalainen}
\email{panu.ruotsalainen@jyu.fi}
\altaffiliation[Present address: ]{University of Jyvaskyla, Department of Physics, P. O. Box 35, FI-40014 University of Jyvaskyla, Finland}
\affiliation{TRIUMF, 4004 Wesbrook Mall, Vancouver, BC, V6T 2A3, Canada}

\author{J.~Henderson}
\altaffiliation[Present address: ]{Lawrence Livermore National Laboratory, 7000 East Ave, Livermore, CA 94550, USA}
\affiliation{TRIUMF, 4004 Wesbrook Mall, Vancouver, BC, V6T 2A3, Canada}

\author{G.~Hackman}
\affiliation{TRIUMF, 4004 Wesbrook Mall, Vancouver, BC, V6T 2A3, Canada}

\author{G.H.~Sargsyan}
\affiliation{Department of Physics and Astronomy, Louisiana State University, Baton Rouge, LA 70803, USA}

\author{K.~D.~Launey}
\affiliation{Department of Physics and Astronomy, Louisiana State University, Baton Rouge, LA 70803, USA}

\author{A.~Saxena}
\affiliation{Department of Physics, Indian Institute of Technology Roorkee, Roorkee 247 667, India}

\author{P.~C.~Srivastava}
\affiliation{Department of Physics, Indian Institute of Technology Roorkee, Roorkee 247 667, India} 

\author{S.~R.~Stroberg}
\altaffiliation[Present address: ]{Department of Physics, University of Washington, Seattle, WA  98195, USA}
\affiliation{TRIUMF, 4004 Wesbrook Mall, Vancouver, BC, V6T 2A3, Canada}

\author{T.~Grahn}
\affiliation{University of Jyvaskyla, Department of Physics, P. O. Box 35, FI-40014 University of Jyvaskyla, Finland}

\author{J.~Pakarinen}
\affiliation{University of Jyvaskyla, Department of Physics, P. O. Box 35, FI-40014 University of Jyvaskyla, Finland}

\author{G.~C.~Ball}
\affiliation{TRIUMF, 4004 Wesbrook Mall, Vancouver, BC, V6T 2A3, Canada}

\author{R.~Julin}
\affiliation{University of Jyvaskyla, Department of Physics, P. O. Box 35, FI-40014 University of Jyvaskyla, Finland}

\author{P.~T.~Greenlees}
\affiliation{University of Jyvaskyla, Department of Physics, P. O. Box 35, FI-40014 University of Jyvaskyla, Finland}

\author{J.~Smallcombe}
\affiliation{TRIUMF, 4004 Wesbrook Mall, Vancouver, BC, V6T 2A3, Canada}

\author{C.~Andreoiu}
\affiliation{Department of Chemistry, Simon Fraser University, Burnaby, BC, V5A 1S6, Canada}

\author{N.~Bernier}
\affiliation{TRIUMF, 4004 Wesbrook Mall, Vancouver, BC, V6T 2A3, Canada}
\affiliation{Department of Physics and Astronomy, University of British Columbia, Vancouver, BC, V6T 1Z4, Canada}

\author{M.~Bowry}
\affiliation{TRIUMF, 4004 Wesbrook Mall, Vancouver, BC, V6T 2A3, Canada}

\author{M.~Buckner}
\affiliation{Lawrence Livermore National Laboratory, 7000 East Ave, Livermore, CA 94550, USA}

\author{R.~Caballero-Folch}
\affiliation{TRIUMF, 4004 Wesbrook Mall, Vancouver, BC, V6T 2A3, Canada}

\author{A.~Chester}
\altaffiliation[Present address: ]{TRIUMF, 4004 Wesbrook Mall, Vancouver, BC, V6T 2A3, Canada}
\affiliation{Department of Chemistry, Simon Fraser University, Burnaby, BC, V5A 1S6, Canada}

\author{S.~Cruz}
\affiliation{TRIUMF, 4004 Wesbrook Mall, Vancouver, BC, V6T 2A3, Canada}
\affiliation{Department of Physics and Astronomy, University of British Columbia, Vancouver, BC, V6T 1Z4, Canada}

\author{L.~J.~Evitts}
\affiliation{TRIUMF, 4004 Wesbrook Mall, Vancouver, BC, V6T 2A3, Canada}
\affiliation{Department of Physics, University of Surrey, Guildford, Surrey, GU2 7XH, UK}

\author{R.~Frederick}
\affiliation{TRIUMF, 4004 Wesbrook Mall, Vancouver, BC, V6T 2A3, Canada}

\author{A.~B.~Garnsworthy}
\affiliation{TRIUMF, 4004 Wesbrook Mall, Vancouver, BC, V6T 2A3, Canada}

\author{M.~Holl}
\affiliation{Astronomy and Physics Department, Saint Mary's University, Halifax, Nova Scotia, B3H 3C3, Canada}
\affiliation{TRIUMF, 4004 Wesbrook Mall, Vancouver, BC, V6T 2A3, Canada}

\author{A.~Kurkjian}
\altaffiliation[Present address: ]{Department of Physics, Simon Fraser University, Burnaby, BC, V5A 1S6, Canada}
\affiliation{TRIUMF, 4004 Wesbrook Mall, Vancouver, BC, V6T 2A3, Canada}

\author{D.~Kisliuk}
\affiliation{Department of Physics, University of Guelph, Guelph, ON, N1G 2W1, Canada}

\author{K.~G.~Leach}
\affiliation{Department of Physics, Colorado School of Mines, Golden, CO 80401, USA}

\author{E.~McGee}
\affiliation{Department of Physics, University of Guelph, Guelph, ON, N1G 2W1, Canada}

\author{J.~Measures}
\affiliation{TRIUMF, 4004 Wesbrook Mall, Vancouver, BC, V6T 2A3, Canada}
\affiliation{Department of Physics, University of Surrey, Guildford, Surrey, GU2 7XH, UK}

\author{D.~M\"ucher}
\affiliation{Department of Physics, University of Guelph, Guelph, ON, N1G 2W1, Canada}

\author{J.~Park}
\altaffiliation[Present address: ]{Department of Physics, Lund University, 22100 Lund, Sweden}
\affiliation{TRIUMF, 4004 Wesbrook Mall, Vancouver, BC, V6T 2A3, Canada}
\affiliation{Department of Physics and Astronomy, University of British Columbia, Vancouver, BC, V6T 1Z4, Canada}

\author{F.~Sarazin}
\affiliation{Department of Physics, Colorado School of Mines, Golden, CO 80401, USA}

\author{J.~K.~Smith}
\altaffiliation[Present address: ]{Department of Physics, Pierce College Puyallup, Washington, WA 98374, USA}
\affiliation{TRIUMF, 4004 Wesbrook Mall, Vancouver, BC, V6T 2A3, Canada}

\author{D.~Southall}
\altaffiliation[Present address: ]{Department of Physics, University of Chicago, Chicago, Illinois 60637, USA}
\affiliation{TRIUMF, 4004 Wesbrook Mall, Vancouver, BC, V6T 2A3, Canada}

\author{K.~Starosta}
\affiliation{Department of Chemistry, Simon Fraser University, Burnaby, BC, V5A 1S6, Canada}

\author{C.~E.~Svensson}
\affiliation{Department of Physics, University of Guelph, Guelph, ON, N1G 2W1, Canada}

\author{K.~Whitmore}
\affiliation{Department of Chemistry, Simon Fraser University, Burnaby, BC, V5A 1S6, Canada}

\author{M.~Williams}
\affiliation{TRIUMF, 4004 Wesbrook Mall, Vancouver, BC, V6T 2A3, Canada}

\author{C.~Y.~Wu}
\affiliation{Lawrence Livermore National Laboratory, 7000 East Ave, Livermore, CA 94550, USA}

\date{\today}

\begin{abstract}
The $T_z$~=~$-\frac{3}{2}$ nucleus ${}^{21}$Mg has been studied by Coulomb excitation on ${}^{196}$Pt and ${}^{110}$Pd targets. A 205.6(1)-keV $\gamma$-ray transition resulting from the Coulomb excitation of the $\frac{5}{2}^+$ ground state to the first excited $\frac{1}{2}^+$ state in ${}^{21}$Mg was observed for the first time. Coulomb excitation cross-section measurements with both targets and a measurement of the half-life of the $\frac{1}{2}^+$ state yield an adopted value of $B(E2;\frac{5}{2}^+\rightarrow\frac{1}{2}^+)$~=~13.3(4)~W.u. A new excited state at 1672(1)~keV with tentative $\frac{9}{2}^+$ assignment was also identified in ${}^{21}$Mg. This work demonstrates large difference of the $B(E2;\frac{5}{2}^+\rightarrow\frac{1}{2}^+)$ values between $T$~=~$\frac{3}{2}$, $A$~=~21 mirror nuclei. The difference is investigated in the shell-model framework employing both isospin conserving and breaking USD interactions and using modern \textsl{ab initio} nuclear structure calculations, which have recently become applicable in the $sd$ shell.
\end{abstract}

\maketitle

\section{\label{sec:level1}Introduction}
Nuclei around the \textit{N}~=~\textit{Z} line serve as a laboratory to investigate the level to which isospin symmetry is conserved in nature. Traditionally isospin symmetry and its breaking have been investigated by comparing the energies of excited states in mirror nuclei or their masses~\cite{bentley}. In order to further the understanding of isospin symmetry breaking effects and develop the existing nuclear models, a range of spectroscopic data is required, including $B(E2)$ values, in addition to level energies and nuclear masses. Nuclear structure studies in the $sd$ shell are particularly interesting since this region is accessible by nuclear theory through phenomenological and \textsl{ab initio} methods.\par

The phenomenological isospin symmetric USD interaction~\cite{brown} was successful in reproducing experimental data, but required additional corrections to reproduce the mirror energy difference (MED) systematics of the 2$^+$ states in $A$~=~18-36, $T$~=~1,2 nuclei~\cite{door}. The main modification of the USD interaction was the use of experimental single-particle energies derived from the $A$~=~17, $T$~=~$\frac{1}{2}$ mirror pair, which implicitly introduce isospin symmetry breaking since the excitation energies in ${}^{17}$O and ${}^{17}$F are likely influenced by the Thomas-Ehrman shift~\cite{ehrman,thomas} and other Coulomb effects. Additional corrections to the calculation were performed separately for the nuclei lying in the lower ($A$~=~18-28) and higher ($A$~=~28-36) $sd$ shells~\cite{door} yielding a very good agreement with experimental MED.\par

Subsequently, the modified USD interaction (USD$^m_{2,3}$) was applied to calculate both MED and $B(E2)$ values in $T$~=~1,$\frac{3}{2}$,2 $sd$-shell mirror pairs~\cite{wendt}. The MED values in these systems are experimentally well known. Experimental $B(E2;0^+_1~\rightarrow~2^+_1)$ and $B(E2;\frac{5}{2}^+_1~\rightarrow~\frac{1}{2}^+_1)$ values for $T_z$~=~$+$1,$+$2 and $T_z$~=~$+\frac{3}{2}$ nuclei, respectively, are also available at or near the valley of stability. However, for neutron-deficient $T_z$~=~$-\frac{3}{2}$,$-$2 nuclei the available experimental data are scarce. For example, information on the $B(E2)$ values in $T_z$~=~$-\frac{3}{2}$ nuclei was limited to ${}^{33}$Ar~\cite{wendt} prior to the present work.\par     

The MED values in $A$~=~19-37, $T$~=~$\frac{3}{2}$ mirror pairs have been reasonably well reproduced by the USD$^m_{2,3}$ interaction. The same is also true for the $B(E2)$ values in $T_z$~=~$\pm$1, $+\frac{3}{2}$ and $\pm$2 nuclei between mass ranges of $A$~=~18-38, 21-37 and 20-36, respectively. The first experimental $B(E2)$ value for the $T_z$~=~$-\frac{3}{2}$ nucleus ${}^{33}$Ar was found to be in excellent agreement with the USD$^m_{2,3}$ prediction. However, it is unclear if the USD$^m_{2,3}$ interaction is actually required to reproduce $B(E2)$ data like it clearly is in the case of MED. Moreover, the USD$^m_{2,3}$ calculation predicted a large difference between $B(E2)$ values in $A$~=~21, $T$~=~$\frac{3}{2}$ mirror nuclei (${}^{21}$Mg/${}^{21}$F)~\cite{wendt}, but it was not quantified what fraction of this difference, if any, had its origin in isospin breaking interactions. \par

\begin{figure}
\includegraphics[scale=0.50]{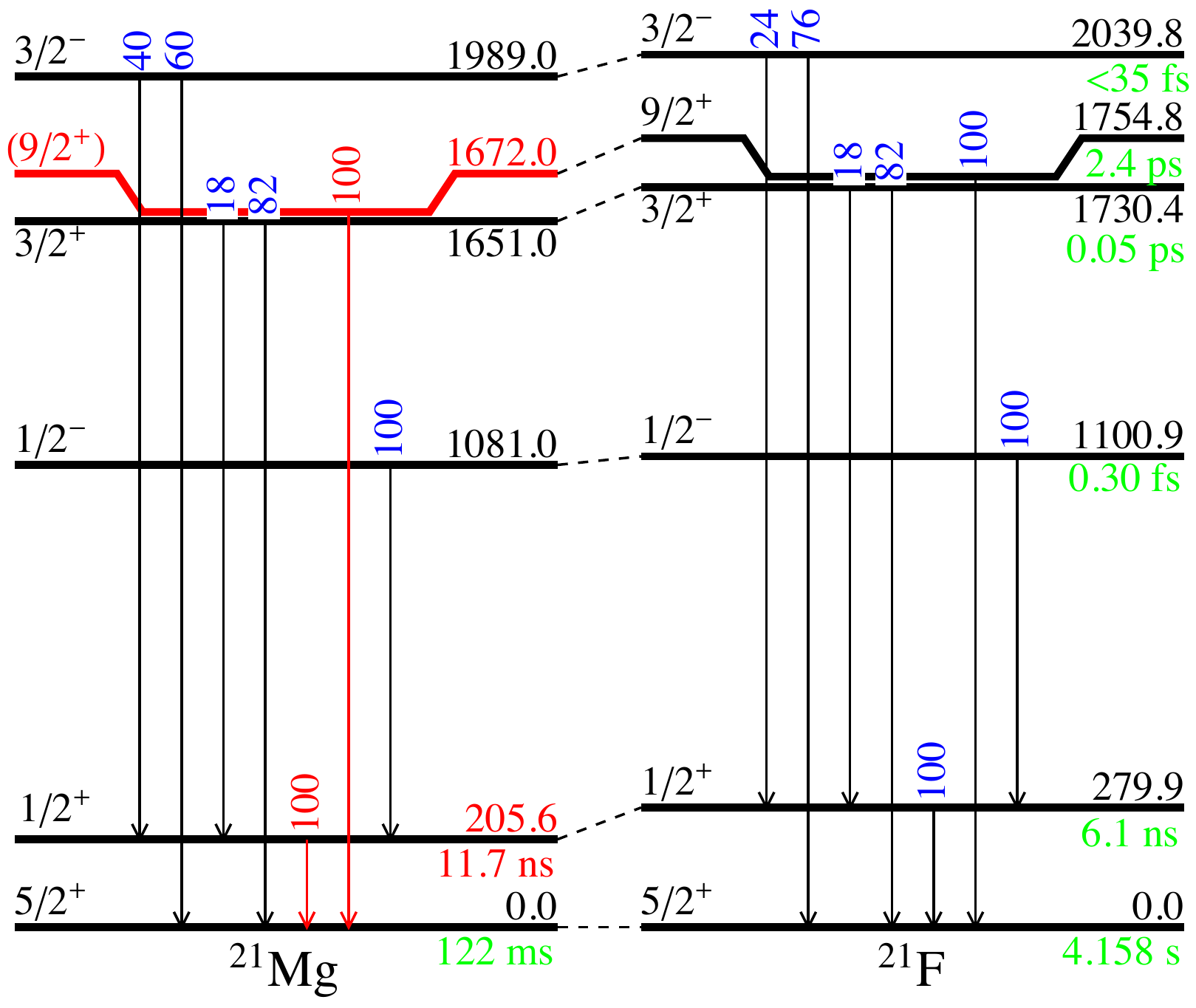}
\colorcaption{\label{fig_scheme} The low-lying level schemes of ${}^{21}$Mg and ${}^{21}$F with known $\gamma$-ray transitions, branching ratios (blue (dark grey) figures) and level half-lives (green (light grey) figures). Data is obtained from Refs.~\cite{firestone,kozub,vonmos}. Spectroscopic information obtained in the present work is indicated in red (gray).}
\end{figure}

The low-lying level schemes of ${}^{21}$Mg and ${}^{21}$F with available spectroscopic information, including new data from the present work, are presented in Fig.~\ref{fig_scheme}. Prior to this work, no $\gamma$-ray transition from the lowest-lying state had been observed. In the present work the $B(E2;\frac{5}{2}^+~\rightarrow~\frac{1}{2}^+)$ value in ${}^{21}$Mg is extracted for the first time using both Coulomb excitation and electronic timing. The obtained $B(E2)$ value together with other available $B(E2)$ data for $T$~=~$\frac{3}{2}$ mirror nuclei are compared to the USD$^m_{2,3}$ prediction, but also to the isospin conserving USDB calculation. Aim is to investigate the importance of the isospin symmetry breaking modifications of the USD interaction specifically on $B(E2)$ values. Predictions obtained from modern \textsl{ab initio} calculations that include isospin symmetry breaking at the nucleon-nucleon interaction level will also be compared to the available experimental $B(E2)$ data.

\section{\label{sec:level2}Experimental setup and details}
The experiment was performed at the TRIUMF~-~ISAC-II facility in Vancouver, Canada. A proton beam with 70-$\mu$A intensity, accelerated with TRIUMF's main cyclotron to 500-MeV energy, impinged on a SiC target~\cite{dombsky}. Spallation reaction products were ionized using the TRIUMF Resonant Ionization Laser Ion Source (TRILIS)~\cite{bricault} to enhance the $^{21}$Mg yield with respect to the three orders of magnitude higher $^{21}$Na yield. The $^{21}$Na contamination was heavily suppressed by mass selection in the ISAC mass separator after which the ions were injected to the ISAC and ISAC-II linear accelerator chain. The post-accelerated $^{21}$Mg ions were delivered to the TIGRESS~\cite{tigress} experimental station with two different beam energies; 95~MeV was used with a 2.93-mg/cm$^2$ thick ${}^{196}$Pt target enriched to 94.6~\%, while 67~MeV was used with a 2.94-mg/cm$^2$ thick ${}^{110}$Pd foil enriched to 97.6~\%. Data were collected with the ${}^{196}$Pt and ${}^{110}$Pd targets for $\sim$66~h and $\sim$24~h, respectively. The average ${}^{21}$Mg intensity at the TIGRESS target position was approximately 5~$\times$~10$^5$~particles/s. The beam composition was monitored by employing a Bragg detector~\cite{tbragg}. The ${}^{21}$Na contamination was found to vary between 16-19~\% of the total beam intensity.\par

The ${}^{21}$Mg ions were Coulomb excited on the ${}^{196}$Pt and ${}^{110}$Pd targets housed within the BAMBINO chamber located at the center of the TIGRESS~\cite{tigress} germanium-detector array. For the ${}^{196}$Pt target, 95~MeV is the highest safe bombarding energy for which the Coulomb excitation process is still purely electromagnetic at all angles according to the Cline criterion~\cite{cline}. For the ${}^{110}$Pd target, 67-MeV energy is safe up to the center-of-mass angle 145$^\circ$. Coulomb excitation induced $\gamma$ rays from the beam and target nuclei were detected with 14 HPGe clover detectors each equipped with BGO and CsI(Tl) Compton suppressors. The TIGRESS detectors were arranged in the high-efficiency configuration providing absolute photopeak efficiency of 11.3(7)~\% at 1.3~MeV. Scattered ${}^{21}$Mg projectiles were detected with the BAMBINO array consisting of two 150-$\mu$m thick annular Micron S3-type silicon detectors~\cite{hurst,kwan,micron_manual} located 30~mm up- and downstream from the target position. The BAMBINO detectors cover laboratory $\theta$ angles between 20.1$^\circ$-49.9$^\circ$ and 130.6$^\circ$-159.9$^\circ$.\par

The TIGRESS digital data acquisition system~\cite{tigress} was used to acquire data in particle singles and particle-$\gamma$ coincidence trigger modes. Preamplifier waveforms (traces) from all detectors were recorded on an event-by-event basis. Traces were fitted offline to improve the electronic timing resolution~\cite{rizwan}. A linear fit is made to the baseline while quadratic and linear fits are applied to the rising edges of the Ge and Si traces, respectively. Time of a radiation event is extracted with $\sim$1~ns accuracy from the intersection of the two fits. Depending on the $\gamma$-ray energy, tens of ns timing resolution for the prompt Ge-Si coincidences was obtained.

 
\section{\label{sec:level3}Analysis and Results}
\begin{figure}
\includegraphics[scale=0.205]{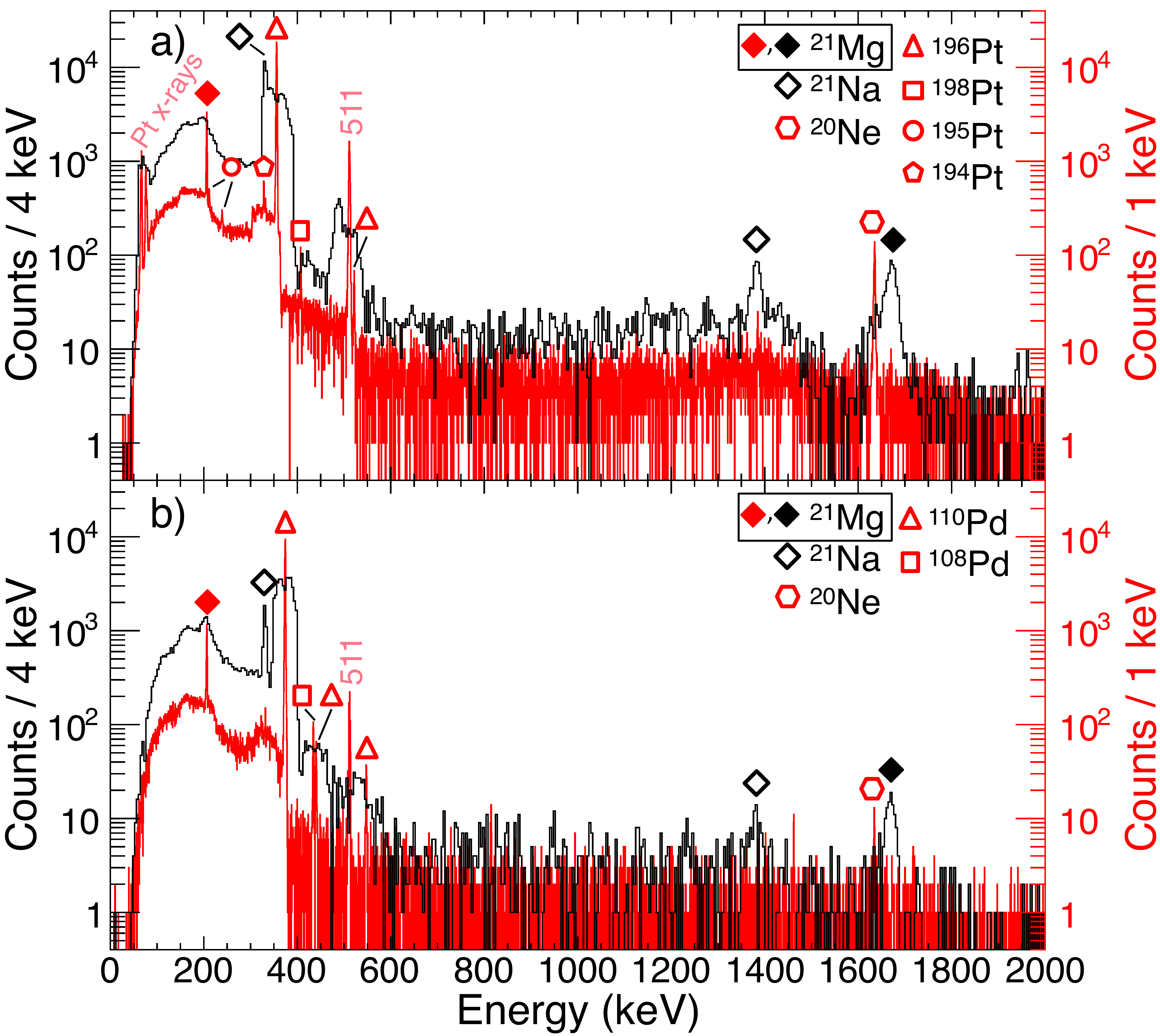}
\colorcaption{\label{fig1} Energy spectra of $\gamma$ rays gated by the $A$~=~21 recoils detected in the downstream Si detector and scattered from a) ${}^{196}$Pt and b) ${}^{110}$Pd target. The black curves are with the Doppler correction, while the red (gray) curves are without it. The peaks at 205.6(1)~keV (solid red (gray) diamonds) and 1672(1)~keV (solid black diamonds) are the first direct observations of the ${\frac{1}{2}}^+\rightarrow{\frac{5}{2}}^+$ and ${\frac{9}{2}}^+\rightarrow{\frac{5}{2}}^+$ $\gamma$-ray transitions in ${}^{21}$Mg, respectively.}
\end{figure}

The $\gamma$-ray energy spectra with the Doppler correction (black curve) and without it (red (gray) curve) observed in coincidence with the $A$~=~21 (${}^{21}$Mg and ${}^{21}$Na) projectiles scattered downstream from the ${}^{196}$Pt and ${}^{110}$Pd targets are presented in Fig.~\ref{fig1}~a) and b), respectively. Previous studies have identified a state at $\sim$200~keV in ${}^{21}$Mg, but $\gamma$-ray transitions from this state were not observed~\cite{kubono,diget}. Ref.\cite{kubono} suggests $\frac{1}{2}^+$ assignment for this state based on the measured angular distributions of three-particle transfer. The ${}^{21}$Mg ground-state spin is measured to be $J=\frac{5}{2}$~\cite{kramer} and comparison with ${}^{21}$F suggests positive parity. The non-observation of the $\frac{1}{2}^+\rightarrow\frac{5}{2}^+$ $\gamma$-ray transition in Ref.~\cite{diget} was attributed to the isomeric nature of the $\frac{1}{2}^+$ state. The analogue $\frac{1}{2}^+$ state in ${}^{21}$F has a half-life of t$_{1/2}$~=~6.1(2)~ns~\cite{warb}.\par

In the present work, a $\gamma$-ray line was observed at 205.6(1)~keV labeled with the red (gray) solid diamonds in Fig.~\ref{fig1}. This transition must originate from ${}^{21}$Mg since it was not observed when the TRILIS lasers were blocked. The measured energy is in agreement with the previously measured $\frac{1}{2}^+$ state energies of 208(10)~keV~\cite{kubono} and 201(4)~keV~\cite{diget}. Since the 205.6(1)-keV transition was observed without employing the Doppler correction, the half-life of the initial state has to be sufficiently long for the excited projectile to reach the S3 detector, where the $\gamma$-ray emission takes place. Consequently, the observed 205.6(1)-keV line signifies the first direct observation of the $\frac{1}{2}^+$~$\rightarrow$~$\frac{5}{2}^+$ $\gamma$-ray transition in ${}^{21}$Mg. The other $\gamma$-ray lines in Fig.~\ref{fig1} labeled with open red (gray) symbols arise from the Coulomb excitation of the target nuclei and from ${}^{20}$Ne, which is populated in the $\beta$-delayed proton decay of ${}^{21}$Mg. \par

Figure~\ref{fig1} shows also the $\gamma$-ray energies, which have been Doppler corrected on an event-by-event basis for ${}^{21}$Mg and ${}^{21}$Na using the position information obtained from the Si and Ge detectors. This results in an energy resolution of 20~keV at 1.384~MeV. The two lines at 332.0(3)~keV and 1384(1)~keV labeled with open black diamonds correspond to the $\frac{5}{2}^+\rightarrow\frac{3}{2}^+$ and $\frac{7}{2}^+\rightarrow\frac{5}{2}^+$ transitions in ${}^{21}$Na, respectively. The 1672(1)-keV line labeled in Fig.~\ref{fig1} with the solid black diamond is assigned to originate from ${}^{21}$Mg because there are no corresponding transitions in the target nuclei or in ${}^{21}$Na.\par

As shown in Fig.~\ref{fig_scheme}, there are $\frac{3}{2}^+$ and $\frac{9}{2}^+$ states at 1730~keV and 1755~keV, respectively, in ${}^{21}$F with only 25-keV energy difference~\cite{kozub,vonmos}. A $\frac{3}{2}^+$ state at 1651(10)~keV has been previously identified in ${}^{21}$Mg~\cite{kubono,diget}. This state is the isobaric analogue of the 1730-keV state in ${}^{21}$F with identical decay branching ratios. It seems likely that the newly observed 1672(1)-keV $\gamma$-ray transition originates from a state in ${}^{21}$Mg, which is the isobaric analogue of the $\frac{9}{2}^+$ state in ${}^{21}$F. The new state at 1672(1)~keV lies 21~keV above the previously identified $\frac{3}{2}^+$ state in good agreement with the mirror nucleus. Consequently, this work demonstrates the first experimental observation of the $\frac{9}{2}^+$ state in ${}^{21}$Mg.\par

In order to extract the $B(E2)$ values in ${}^{21}$Mg, the Coulomb excitation data were divided into six subsets corresponding to six ranges of projectile scattering angles covered by the downstream S3 detector. Data collected with the ${}^{196}$Pt and ${}^{110}$Pd targets were analyzed separately. The intensities of the $\frac{1}{2}^+\rightarrow\frac{5}{2}^+$ and $\frac{9}{2}^+\rightarrow\frac{5}{2}^+$ $\gamma$-ray transitions in ${}^{21}$Mg for each subset of data were extracted and corrected for the detection efficiency and the target impurity~\cite{gosia_methods}. The intensity of the 205.6(1)-keV line was extracted initially from the decays occurring in the downstream S3 detector since the in-flight decay component could not be observed in the Doppler corrected spectra. The intensities of the $\gamma$-ray lines resulting from the target excitations were extracted and corrected for the detection efficiency and the beam impurity~\cite{gosia_methods}. The detection efficiency of TIGRESS was measured at the target position and at the locations of the S3 detectors using ${}^{152}$Eu and ${}^{133}$Ba sources. \par

Corrected ${}^{21}$Mg and target $\gamma$-ray yields were analyzed using the GOSIA2 code~\cite{gosia_methods, gosia_orig, gosia_manual}. The ${}^{21}$Mg matrix elements were fitted relative to the ${}^{196}$Pt and ${}^{110}$Pd target $\gamma$-ray yields. Matrix elements of the low-lying transitions in both targets are known with good precision together with other spectroscopic data~\cite{pt196,pd110}, which allow them to be used as an absolute normalisation for the beam excitations. For ${}^{21}$Mg the two observed states and their matrix elements in addition to a buffer state above the $\frac{9}{2}^+$ state were included in the analyses. The presently measured $\frac{1}{2}^+$ state half-life was not utilized in the GOSIA2 analyses in order to ensure the independence of the analyses.\par

In the GOSIA2 fitting procedure $\langle\frac{1}{2}^+||E2||\frac{5}{2}^+\rangle$ and $\langle\frac{9}{2}^+||E2||\frac{5}{2}^+\rangle$ were scanned simultaneously resulting in a two-dimensional $\chi^2$ surface. The minimum $\chi^2$ value ($\chi^2_{min}$) represents the best fit of the matrix elements to the experimental $\gamma$-ray yields~\cite{gosia_methods}. This analysis was performed iteratively since the $\frac{1}{2}^+$ state decays partly between the target and the S3 detector reducing the true $\gamma$-ray yield. The obtained $\langle\frac{1}{2}^+||E2||\frac{5}{2}^+\rangle$ matrix element from the first (previous) step was employed to compute the half-life of the $\frac{1}{2}^+$ state, which was then used to correct the $\frac{1}{2}^+\rightarrow\frac{5}{2}^+$ $\gamma$-ray transition intensities for in-flight decay losses for the next analysis round. The $\langle\frac{1}{2}^+||E2||\frac{5}{2}^+\rangle$ values converged rapidly after 4 analysis steps for both ${}^{196}$Pt and ${}^{110}$Pd data increasing the non-corrected matrix elements by 4~\% and 3~\%, respectively. \par

\begin{figure}
\includegraphics[scale=0.45]{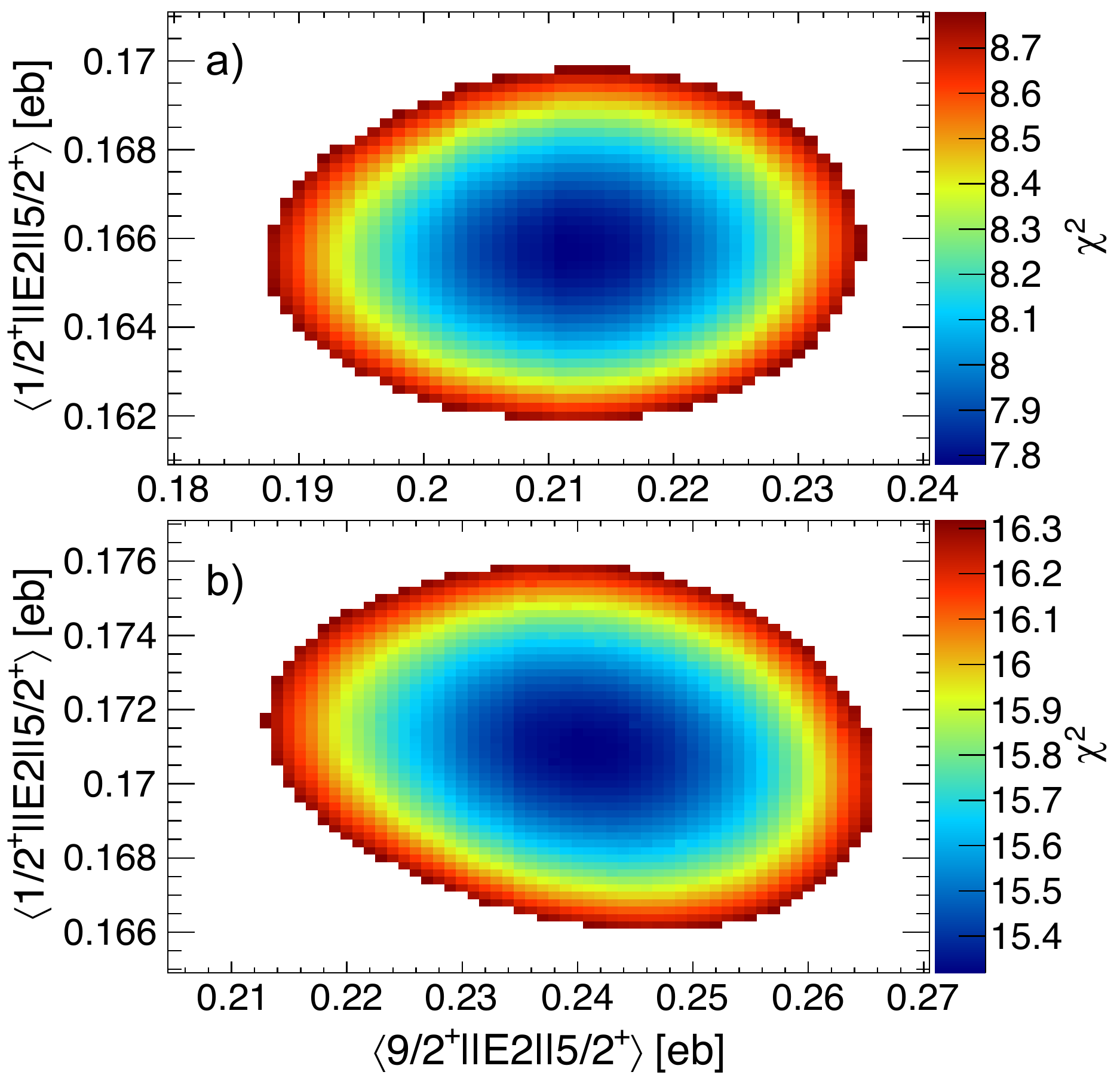}
\colorcaption{\label{fig_maps} Two-dimensional $\chi^2$ surfaces obtained from the GOSIA2 analysis performed with a) ${}^{196}$Pt and b) ${}^{110}$Pd target data with applied $\chi^2 < \chi^2_{min}+1$ criterion.}
\end{figure}

The $\chi^2$ surfaces with applied 1$\sigma$ cuts are shown in Fig.~\ref{fig_maps} a) and b) for the ${}^{196}$Pt and ${}^{110}$Pd target data, respectively, after the convergence was reached. The 1$\sigma$-uncertainty contour is the part of the $\chi^2$ surface for which $\chi^2 < \chi^2_{min}+1$. The uncertainties of the matrix elements are obtained by projecting the 1$\sigma$ contour on the corresponding matrix element axis~\cite{gosia_methods}. Matrix elements and $B(E2)\uparrow$ values with errors are presented in Table~\ref{table1}. The obtained matrix elements from different measurements are in good agreement within uncertainties.\par
	
\begin{table} 
\caption{Matrix elements and $B(E2)\uparrow$ values for ${}^{21}$Mg from the GOSIA2 analysis with ${}^{196}$Pt and ${}^{110}$Pd targets and from the half-life measurement of the $\frac{1}{2}^+$ state. } 
\label{table1}
\begin{tabular*}{0.48\textwidth}{p{0.17\textwidth}p{0.1\textwidth}p{0.1\textwidth}p{0.1\textwidth}}\hline \hline\noalign{\smallskip}
 \hspace{0.0cm} ${}^{21}$Mg & ${}^{196}$Pt target & ${}^{110}$Pd target & from t$_{1/2}$ \\
\noalign{\smallskip}\hline\noalign{\smallskip}
 $\langle\frac{1}{2}^+||E2||\frac{5}{2}^+\rangle$~[eb] & 0.166(4) & 0.171(5) & 0.162(4) \\

\noalign{\smallskip}
$B(E2;\frac{5}{2}^+\rightarrow\frac{1}{2}^+)$ &  &  &  \\
\noalign{\smallskip}
 	\hspace{0.0cm}[W.u.] & 13.3(6) & 14.2(8) & 12.7(6) \\
\noalign{\smallskip}\hline\noalign{\smallskip}
 $\langle\frac{9}{2}^+||E2||\frac{5}{2}^+\rangle$~[eb] & 0.21(2) & 0.24(3) & $-$ \\
\noalign{\smallskip}
$B(E2;\frac{5}{2}^+\rightarrow\frac{9}{2}^+)$ &  &  &  \\
\noalign{\smallskip}
 	\hspace{0.0cm}[W.u.] & 22(5) & 28(7) & $-$ \\
\noalign{\smallskip}\hline \hline
\end{tabular*}
\end{table}

\begin{figure}
\includegraphics[scale=0.42]{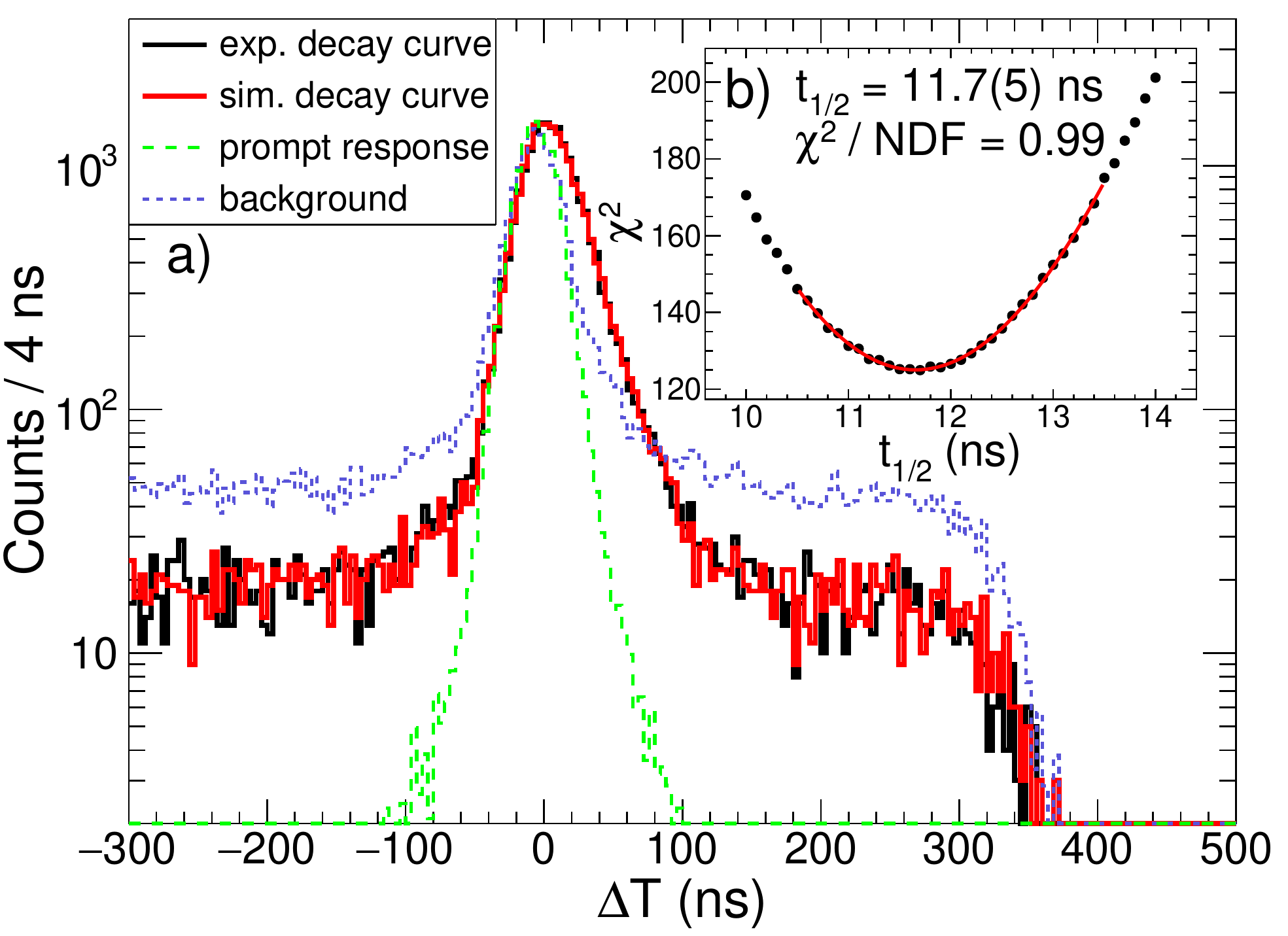}
\colorcaption{\label{fig3} a) Experimental decay curve of the $\frac{1}{2}^+$ state in ${}^{21}$Mg (black) together with the best fit simulated decay curve (red (solid gray)). The prompt time difference (green (dashed gray)) and time random background (violet (short dashed gray)) sampling distributions are also presented. b) The obtained $\chi^2$ values as a function of $t_{1/2}$ of the simulated activity.} 
\end{figure}

The decay curve of the $\frac{1}{2}^+$ state with $\sim$2.2~$\times$~10$^4$ events was obtained from the Ge$-$S3 time difference distribution gating on the 205.6(1)-keV $\gamma$ rays (black line in Fig.~\ref{fig3}~a)). This was then compared to simulated decay curves (red (solid gray)) generated by sampling $\sim$1.3~$\times$~10$^4$ decay events (= area of the 205.6(1)-keV peak) from the experimental prompt response distribution (green (dashed gray)) and $\sim$0.9~$\times$~10$^4$ events from the background distribution (violet (short dashed gray)) with different half-lives. A $\chi^2$ value was computed for each simulated curve. The prompt response was extracted from the Ge$-$S3 time differences by gating on the 356-keV $\gamma$ rays originating from ${}^{196}$Pt, $2^+$ state with $t_{1/2}$~=~34.15(15)~ps~\cite{pt196}. The width of the distribution was further modified as the timing resolution decreases towards lower $\gamma$-ray energies. The background distribution was obtained by setting gates on both sides of the 205.6(1)-keV peak. Minimum $\chi^2$ was found at $t_{1/2}$~=~11.7(5)~ns as shown in Fig.~\ref{fig3}~b).\par

\section{\label{sec:level4}Discussion}
In the present work, the $B(E2;\frac{5}{2}^+\rightarrow\frac{1}{2}^+)$ value in ${}^{21}$Mg is obtained from three independent measurements $-$ from the Coulomb excitation cross section measurements on ${}^{196}$Pt and ${}^{110}$Pd targets and from the half-life measurement of the $\frac{1}{2}^+$ state. From these measurements the adopted value of $B(E2;\frac{5}{2}^+\rightarrow\frac{1}{2}^+)$~=~13.3(4)~W.u. is obtained using the expected value method~\cite{evm} in V.AveLib software~\cite{vavelib_manual}. This result yields the second data point for the $B(E2)$ value systematics of $T_z$~=~$-\frac{3}{2}$ nuclei in the $sd$ shell. A value of $B(E2;\frac{5}{2}^+\rightarrow\frac{9}{2}^+)$~=~25(4)~W.u. is also obtained in the present work.\par

\begin{figure}[t!]
\includegraphics[scale=0.45]{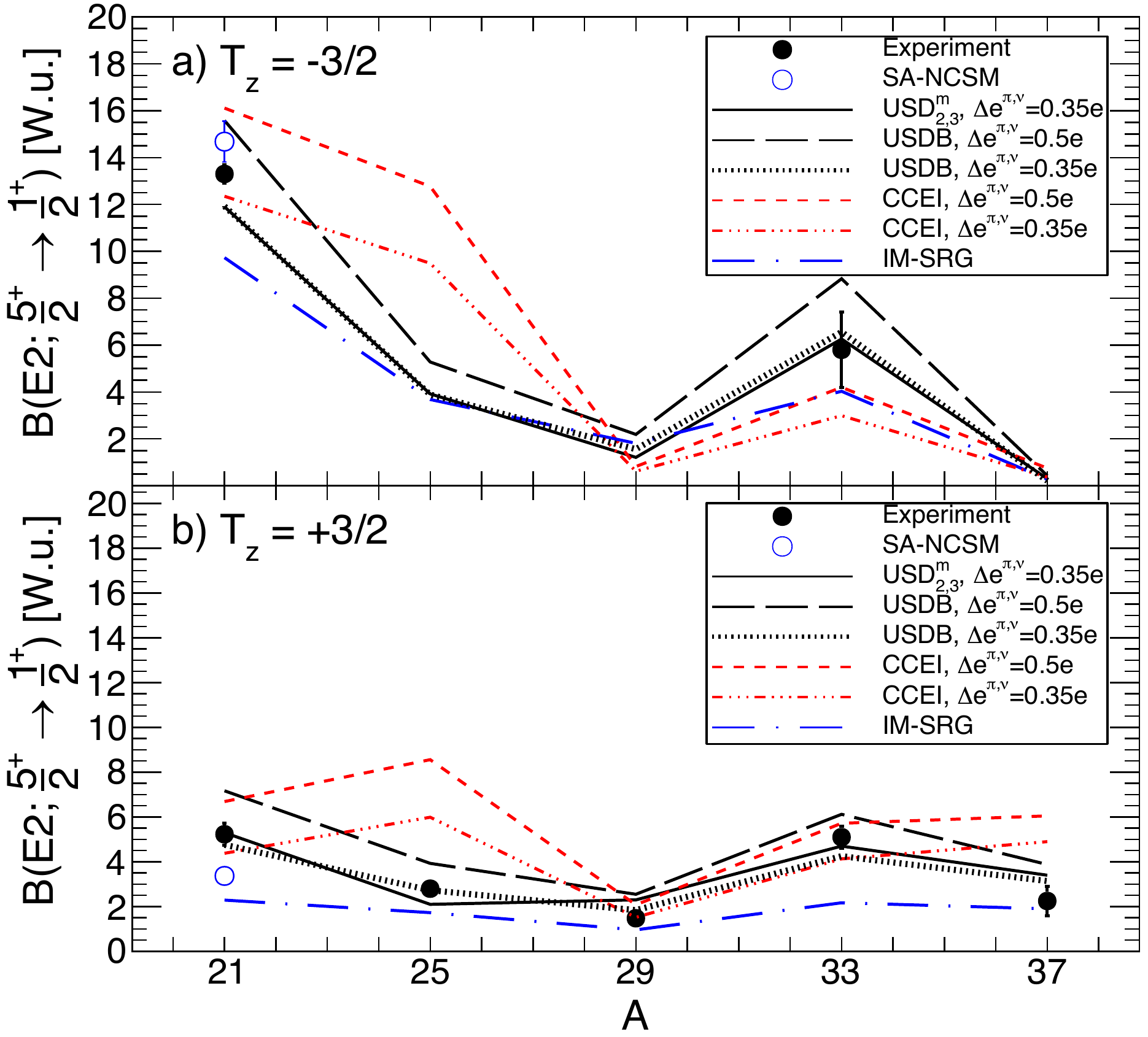}
\colorcaption{\label{fig4} Experimental and theoretical $B(E2;\frac{5}{2}^+\rightarrow\frac{1}{2}^+)$ values for a) $T_z$~=~$-\frac{3}{2}$ and b) $T_z$~=~$+\frac{3}{2}$ mirror nuclei including the new experimental value for ${}^{21}$Mg. Theoretical $B(E2)$ values are obtained from the shell-model calculations using USDB and USD$^m_{2,3}$~\cite{wendt} interactions in addition to the SA-NCSM (only for \textsl{A}~=~21), CCEI and IM-SRG \textsl{ab initio} calculations.} 
\end{figure}

The experimental $B(E2;\frac{5}{2}^+\rightarrow\frac{1}{2}^+)$ data for $T_z$~=~$\pm\frac{3}{2}$ nuclei are compared to various theoretical predictions in Fig.~\ref{fig4}~a) and b). The USD$^m_{2,3}$ calculation (taken from Ref.~\cite{wendt}) with an isoscalar polarization charge of $\Delta e^{\pi,\nu}$~=~0.35~e ($e^{\pi}_{\text{eff}}$~=~1.35~e, $e^{\nu}_{\text{eff}}$~=~0.35~e) is in good agreement with the experimental values. The isospin conserving USDB calculation with $\Delta e^{\pi,\nu}$~=~0.35~e yields similar agreement with experiment. This indicates that the $B(E2)$ values, unlike MED, are largely insensitive to the phenomenological isospin symmetry breaking modifications of the USD interaction introduced in Ref.~\cite{door}. The USDB calculation with $\Delta e^{\pi,\nu}$~=~0.5~e is also shown in Fig.~\ref{fig4} to demonstrate $B(E2)$ values' sensitive reliance on the effective charges. The $B(E2)$ values of the \textsl{A}~=~21 mirror pair were further investigated with the USDB-cdpn interaction~\cite{ormand}, which includes Coulomb and charge-dependent interactions, yielding less than 1~\% increase in the $E2$ strength in comparison to USDB.\par

\textsl{Ab initio} methods have recently become available to study the spectroscopic properties of the $sd$ shell nuclei. In Fig.~\ref{fig4} the experimental $B(E2;\frac{5}{2}^+\rightarrow\frac{1}{2}^+)$ values are compared to the coupled-cluster effective interaction (CCEI)~\cite{jansen2}, the in-medium similarity renormalization group (IM-SRG)~\cite{tsukiyama,bogner,stroberg2}, and the symmetry-adapted no-core shell model (SA-NCSM)~\cite{launey,dytrych} calculations. The CCEI, IM-SRG and SA-NCSM methods have been previously applied to calculate the level energies in $p$ and $sd$ shell nuclei~\cite{stroberg,jansen2,jansen1,launey,dytrych}.\par

In the present work the IM-SRG calculation was performed using the EM 1.8/2.0 chiral interaction~\cite{hebeler} in a harmonic oscillator (HO) basis of $\hbar\omega$~=~20~MeV, including 13 major shells. The CCEI calculation employed a similar interaction~\cite{jansen2}. The IM-SRG calculation uses a consistently transformed $E2$ transition operator~\cite{parzu} and does not incorporate effective charges while the CCEI calculation uses a bare transition operator with phenomenological effective charges. The SA-NCSM calculations, not employing effective charges, were performed using the N2LO$_{\text{opt}}$ chiral potential~\cite{ekstrom} with HO frequency range of $\hbar\omega$~=~10$-$20~MeV in a model space of 5 to 13 major shells and three symmetry-based model space selections. For each of these selections, calculations were performed with increasing number of shells to ensure convergence. The results are reported for $\hbar\omega$~=~15~MeV and 13 major shells, while the quoted uncertainties arise from the variation of the $B(E2)$ values with respect to the number of shells and the value of $\hbar\omega$ used in the calculation. Isospin symmetry breaking is included in the IM-SRG, CCEI and SA-NCSM approaches at the level of the chiral interaction. The interactions include the Coulomb force and the smaller non-Coulomb effects due to the different pion masses.\par

The CCEI calculation is found to agree better with experiment with $\Delta e^{\pi,\nu}$~=~0.35~e and it reproduces the experimental $B(E2)$ values at \textsl{A}~=~21 correctly as shown in Fig.~\ref{fig4}. The CCEI results deviate from the other models at \textsl{A}~=~25 since CCEI favours different dominant configurations for the $\frac{1}{2}^+$ states in ${}^{25}$Si and ${}^{25}$Na.\par

The IM-SRG calculation underpredicts the $E2$ strength for the majority of $T$~=~$\frac{3}{2}$ nuclei. The same has been observed with $T_z$~=~$\pm$1, $sd$ shell mirror pairs~\cite{jack}, but the discrepancy was found to be much larger than observed here. The improved agreement achieved here for the $B(E2;\frac{5}{2}^+\rightarrow\frac{1}{2}^+)$ values might result from the $\frac{1}{2}^+$ state configurations, which are likely dominated by single-particle excitations. In particular, IM-SRG is in good agreement with the USD$^m_{2,3}$ and USDB predictions at \textsl{A}~=~25 and \textsl{A}~=~29 where the \textsl{Z, N}~=~14, 16 subshell closures are likely to further suppress collectivity. Nevertheless, the trend for increasing difference of the $B(E2;\frac{5}{2}^+\rightarrow\frac{1}{2}^+)$ values between \textsl{A}~=~21 mirror nuclei is correctly reproduced. This difference is also obtained in the SA-NCSM calculations, which yield larger values in comparison to IM-SRG, but lower and larger values than measured for ${}^{21}$F and ${}^{21}$Mg, respectively. \par

Under the assumption of isospin symmetry, $B(E2;\frac{5}{2}^+\rightarrow\frac{1}{2}^+)$ values from the \textsl{A}~=~21 mirror pair can be used to calculate experimental and theoretical isoscalar ($M_0$) and isovector ($M_1$) matrix elements according to, e.g., Refs.~\cite{brown_isovector,orce}. This analysis implies that the dominant $M_0$ component is correctly reproduced by SA-NCSM, while the $M_1$ component is overestimated by about 50~\% indicating a larger difference between the associated proton $E2$ matrix elements in comparison to the experimental $M_1$.  Similar analysis with the IM-SRG results reveals that the situation is the opposite - the $M_1$ component is only slightly overestimated while the $M_0$ component is clearly underestimated. Whether these observations arise from the characteristic features of the SA-NCSM and IM-SRG approaches remains an open question. \par

\begin{figure}
\includegraphics[scale=0.43]{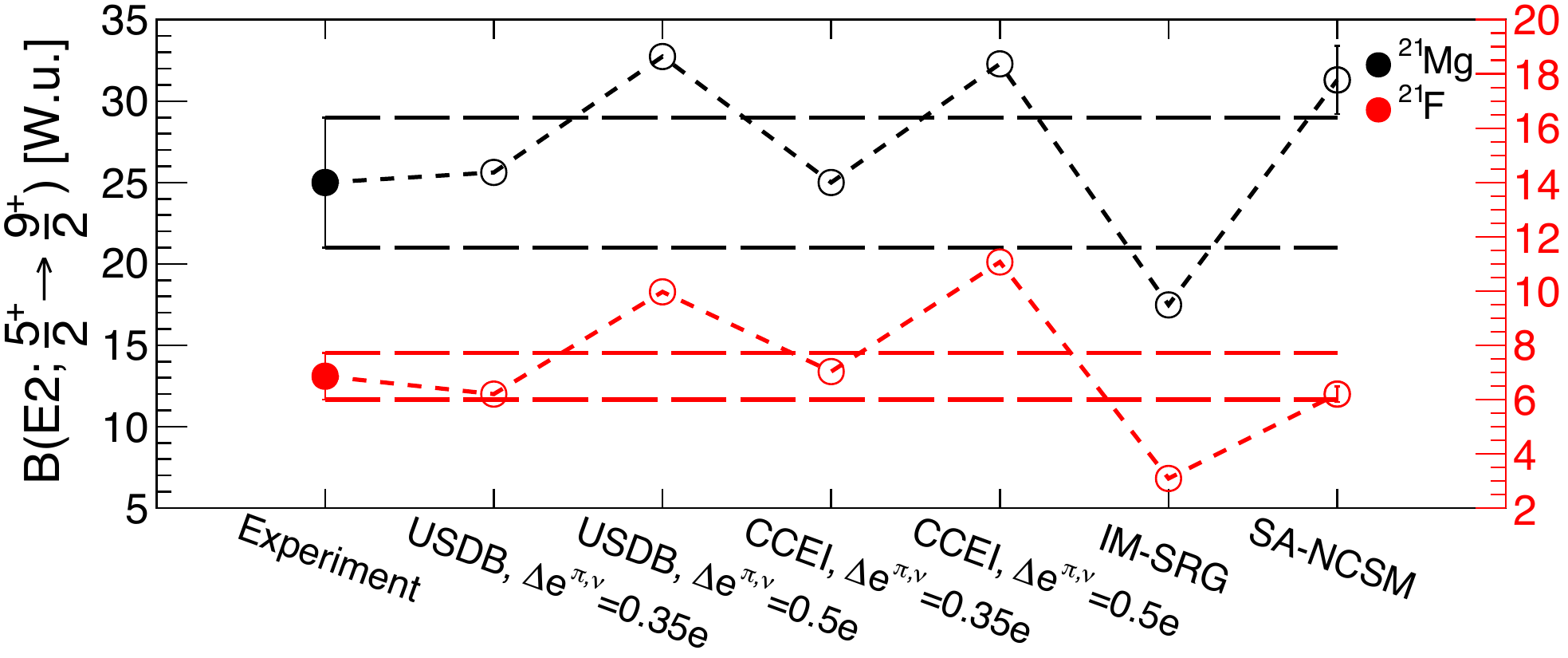}
\colorcaption{\label{fig7} Experimental (solid symbols) and theoretical (open symbols) $B(E2;\frac{5}{2}^+\rightarrow\frac{9}{2}^+)$ values for ${}^{21}$Mg (black) and ${}^{21}$F (red (gray)).} 
\end{figure}

According to USDB calculation a dominant part ($\sim$73~\%) of the $\frac{9}{2}^+$ state in ${}^{21}$Mg is based on $\pi(d_{5/2}^4) \otimes \nu(d_{5/2}^1)$ configuration, which may alternatively be interpreted to arise from a coupling of an odd $d_{5/2}$ neutron to the first excited 2$^+$ state in ${}^{20}$Mg. Figure~\ref{fig7} shows how the different calculations compare with the experimental $B(E2)$ value between the collective $\frac{9}{2}^+$ state and the $\frac{5}{2}^+$ ground state in ${}^{21}$Mg (and ${}^{21}$F). The USDB and CCEI approaches reproduce well the experimental $B(E2)$ values for both nuclei with $\Delta e^{\pi,\nu}$~=~0.35~e. The SA-NCSM calculation lies close to the experimental value in ${}^{21}$Mg, given the quoted uncertainties, while the IM-SRG calculation underpredicts the experimental value by 30~\%.\par


\section{\label{sec:level5}Summary}
The $T_z$~=~$-\frac{3}{2}$ nucleus ${}^{21}$Mg was studied in Coulomb excitation enabling the first direct observations of the $\frac{1}{2}^+\rightarrow\frac{5}{2}^+$ and $\frac{9}{2}^+\rightarrow\frac{5}{2}^+$ $\gamma$-ray transitions. The $B(E2;\frac{5}{2}^+\rightarrow\frac{1}{2}^+)$ and $B(E2;\frac{5}{2}^+\rightarrow\frac{9}{2}^+)$ values were measured and the results are compared to shell-model and \textsl{ab initio} nuclear structure calculations. The $B(E2;\frac{5}{2}^+\rightarrow\frac{1}{2}^+)$ value in ${}^{21}$Mg is found to be more than two times larger than the corresponding value in its mirror nucleus ${}^{21}$F. Shell-model calculations employing modified USD$^m_{2,3}$ and standard USDB interactions reproduce this difference equally well indicating that the associated $B(E2)$ values do not signal significant isospin symmetry breaking. The IM-SRG \textsl{ab initio} approach is found to underpredict both newly measured $B(E2)$ values in ${}^{21}$Mg, while the SA-NCSM \textsl{ab initio} calculations yield a slight overprediction.\par

\begin{acknowledgments}
This work has been supported by the Natural Sciences and Engineering Research Council of Canada (NSERC), The Canada Foundation for Innovation and the British Columbia Knowledge Development Fund. TRIUMF receives federal funding via a contribution agreement through the National Research Council of Canada. The work at LLNL is under contract DE-AC52-07NA27344. The work at JYFL-ACCLAB has been supported by the Academy of Finland under the Finnish Center of Excellence Programme (2012-2017). The work at Colorado School of Mines has been supported by the U.S. Department of Energy under Grant No. DE-SC0017649. This work has been partly supported by the U.S. National Science Foundation (OIA-1738287, ACI-1713690). This work benefitted from computing resources provided by Blue Waters and LSU (\url{www.hpc.lsu.edu}). The Blue Waters sustained-petascale computing project is supported by the National Science Foundation (awards OCI-0725070 and ACI-1238993) and the state of Illinois, and is a joint effort of the University of Illinois at Urbana-Champaign and its National Center for Supercomputing Applications.
\end{acknowledgments}

\bibliographystyle{apsrev}
\bibliography{references}

\end{document}